\documentclass[num-refs]{nbdt-article}

\usepackage{siunitx}
\usepackage[utf8]{inputenc}
\usepackage{amsmath}
\usepackage{amsfonts}
\usepackage{amssymb}
\usepackage{bm}
\usepackage{physics}
\usepackage{pgf}
\usepackage{tikz}
\usepackage{todonotes}
\usetikzlibrary{arrows,automata}
\usepackage{mathrsfs}  
\usepackage{lipsum}
\usepackage{float}
\usepackage[T1]{fontenc}
\usepackage{authblk}
\usepackage{hyperref}
\usepackage{ulem}
\usepackage{booktabs,threeparttable,caption}
\DeclareCaptionLabelFormat{AppendixTables}{A.#2}
\DeclareFontFamily{OT1}{pzc}{}
\DeclareFontShape{OT1}{pzc}{m}{it}{<-> s * [1.10] pzcmi7t}{}
\DeclareMathAlphabet{\mathpzc}{OT1}{pzc}{m}{it}
\usepackage{ccfonts}

\papertype{Original Article}
\paperfield{Neural Data Analysis/Theory}

\title{Frontal effective connectivity increases with task demands and time on task: a Dynamic Causal Model of electrocorticogram in macaque monkeys}

\author[1]{Katharina Wegner}
\author[2]{Charles R.E. Wilson}
\author[2]{Emmanuel Procyk}
\author[3]{Karl J. Friston}
\author[1,3,4]{Frederik Van de Steen}
\author[5,6\authfn{1}]{Dimitris A. Pinotsis}
\author[1\authfn{1}]{Daniele Marinazzo}

\affil[1]{Department of Data Analysis, Ghent University, Belgium}
\affil[2]{Université Claude Bernard Lyon 1, Inserm, Stem Cell and Brain Research Institute U1208, France}
\affil[3]{The Wellcome Centre for Human Neuroimaging, London, United Kingdom}
\affil[4]{AIMS laboratory, Vrije Universiteit Brussel, Brussels, Belgium}
\affil[5]{Centre for Mathematical Neuroscience and Psychology and Department of Psychology, City - University of London, London, United Kingdom}
\affil[6]{The Picower Institute for Learning \& Memory and Department of Brain and Cognitive Sciences, Massachusetts Institute of Technology, Cambridge, MA, 02139, USA}

\contrib[\authfn{1}]{Equal contribution as senior authors.}

\corraddress{Daniele Marinazzo PhD, Department of Data Analysis, Ghent University, Ghent, B9000, Belgium}
\corremail{daniele.marinazzo@ugent.be}

\runningauthor{Wegner et al.}

\begin{document}

\maketitle

\begin{abstract}
We apply Dynamic Causal Models to electrocorticogram recordings from two macaque monkeys performing a problem-solving task that engages working memory, and induces time-on-task effects. We thus provide a computational account of changes in effective connectivity within two regions of the fronto-parietal network, the dorsolateral prefrontal cortex and the pre-supplementary motor area.
We find that forward connections between the two regions increased in strength when task demands increased, and as the experimental session progressed. Similarities in the effects of task demands and time on task allow us to interpret changes in frontal connectivity in terms of increased attentional effort allocation that compensates cognitive fatigue.

\end{abstract}

\section{Introduction}
Engaging in a cognitive task is costly \cite{botvinick_effort_2009,apps_role_2015,shenhav_toward_2017}. Cognitive effort-at least in some attempts to quantify an admittedly vague and wide concept, and as we will interpret it in this study-is the deployment of control that mitigates these costs and helps maintain cognitive performance levels \cite{kahneman_attention_1973,hockey_compensatory_1997}. How much cognitive effort is required to successfully execute a task depends on task demands \cite{coull_fronto-parietal_1996,kitzbichler_cognitive_2011,vassena_overlapping_2014,milyavskaya_reward_2019} or on the duration of task engagement \cite{christensen-szalanski_further_1980,umemoto_electrophysiological_2019}.

These two aspects of cognitive effort—task difficulty and time on task—have both been associated with the frontal beta rhythm in macaque monkeys \cite{stoll_effects_2016,wilson_prefrontal_2016}. In these studies, frontal beta oscillations increased both when cognitive control demands were elevated or as a consequence of sustained task engagement. Given that both aspects of cognitive effort have a common neural correlate, the present study sets out to investigate whether task demands and time on task also share a common modulation of frontal functional connections.

Cognitive functions rely on neural processing within a set of brain regions encompassing the prefrontal, anterior cingulate and parietal regions, known as the frontoparietal network (FPN) \cite{cabeza_imaging_2000, duncan_multiple-demand_2010,spreng_default_2010,vincent_evidence_2008}. Activity within this network increases when the successful execution of an ongoing task requires sustained attention \cite{langner_sustaining_2013}, maintenance of information in working memory \cite{cabeza_imaging_2000,linden_cortical_2003,gazzaley_functional_2004,buschman_top-down_2007,harding_effective_2015} or overriding a prepotent response \cite{harding_effective_2015}, while reduced activation is found during the performance of more habitual tasks \cite{sakai_learning_2002,dosenbach_distinct_2007,landmann_dynamics_2007}. Studies of both monkeys and humans have related beta band oscillations within FPN areas to the maintenance of sensory or task set 
\cite{spitzer_oscillatory_2010,palva_localization_2011,antzoulatos_synchronous_2016} and have shown that FPN beta band power increases with working memory load \cite{palva_localization_2011,kornblith_stimulus_2016}. Of particular interest is the dorsolateral prefrontal cortex (dlPFC): while neurophysiological \cite{goldman-rakic_circuitry_1987,miller_integrative_2001}, neuroimaging \cite{carlson_eye_1998}, and lesion studies \cite{desposito_dependence_1999,ptak_disorganised_2004} converge on the conclusion that the dlPFC is crucial for the maintenance and updating of working memory information, this area also shows increased beta-band activity with time on task \cite{tanaka_neural_2014}. Since other studies have found time-related decreases in dlPFC activity \cite{coull_fronto-parietal_1996,blain_neural_2016}, it is unclear whether the time-related change in neural activity can be attributed to mental fatigue or an increase of cognitive effort that compensates for the effects of fatigue \cite{stoll_effects_2016}.

We propose that the similarity between task demand and time on task effects licence an interpretation of time-on-task-effects as an increase of effort to mitigate the effects of fatigue.
Previous findings have established that cognitive effort can modulate functional and effective connectivity in the FPN. Donner et al. \cite{donner_population_2007} found that beta-band amplitude coherence between the dlPFC and parietal regions increased before correct compared to incorrect responses. Under the assumption that correct responses are accompanied by greater effort compared to incorrect trials, FPN beta coherence may thus be related to cognitive effort. Studies using dynamic causal modeling or structural equation modeling showed that directed connections from the prefrontal to the premotor \cite{koechlin_architecture_2003,koechlin_information_2007,harding_effective_2015} and from parietal regions to the prefrontal areas \cite{ma_working_2012,dima_dynamic_2014, pinotsis_working_2019} changed in strength when task demands increase. Connectivity studies have focused on neural changes related to the quality of performance and demands in tasks requiring cognitive control, but evidence for connectivity changes related to time on task in cognitive control tasks remains scarce. To date, a comparison of how the two aspects of cognitive effort modulate FPN connectivity is therefore lacking.
Two FPN regions—the dlPFC and the pre-supplementary motor area (pre-SMA)—are active during task execution and their activation has been shown to correlate with working memory load \cite{linden_cortical_2003,van_snellenberg_dynamic_2014}. In addition, effective connections from the lateral PFC to the pre-SMA become stronger with increasing task complexity \cite{koechlin_architecture_2003}.

In order to extend these findings, in the present study, we use dynamic causal modeling (DCM) to quantify changes in effective connectivity between the dlPFC and pre-SMA with task demands and time on task. The choice of DCM was further motivated by the evidence that spectral localization of brain rhythms change in individual subjects (see e.g. \cite{kornblith_stimulus_2016}, and that modulations in spectral peaks can be confounded and/or explained by 1/f activity \cite{donoghue2020parameterizing}. The DCM that we use here fits a broad spectral band, and performs a parametrization of the spectrum including aperiodic activity. We use chronic electrocorticographic (ECoG) surface recordings in macaque monkeys that were described in a previous study \cite{stoll_effects_2016}. In the behavioral task, we vary cognitive control demands \cite{procyk_modulation_2006}, while measuring neural and behavioral responses over time. We expected to replicate the finding that bilateral effective connections from the dlPFC to the premotor areas increase in strength with task demands \cite{koechlin_architecture_2003,koechlin_information_2007, badre_cognitive_2008}. Most DCM analysis requires a model comparison to test for the directionality of connections \cite{kiebel_dynamic_2008}. However, since anatomical studies indicate that anatomical connections between the dlPFC and pre-SMA are reciprocal \cite{markov_weighted_2014} and effective connections in both directions have been shown to change with task demands \cite{koechlin_architecture_2003,koechlin_information_2007,ma_working_2012,dima_dynamic_2014,harding_effective_2015}, we only considered a bidirectional dlPFC-pre-SMA model. This assumed that the dlPFC and pre-SMA were at the same hierarchical level. Similar results were found by Pinotsis and colleagues \cite{pinotsis_working_2019} using Bayesian model comparison. In addition to task demand effects, we anticipated time-on-task-effects on connection strength. Furthermore, we expected that both aspects of cognitive effort would show similar modulatory effects. Similarities in the effects of task demands and time on task would licence an interpretation of time-induced changes in frontal connections as increased effort that mitigates cognitive fatigue.

\section{Methods}

\subsection{Ethical Statement}
Ethical permission was provided by “Comit\'{e} \'{E}thique Lyonnais pour les Neurosciences Exp\'{e}rimentales,” CELYNE, C2EA 42, ref: C2EA42-11-11-0402-004. Monkey housing and care was in accordance with European Community Council Directive (2010) and the Weatherall report, The use of non-human primates in research. Laboratory authorization was provided by the "Pr\'{e}fet de la Region Rh\^{o}ne-Alpes" and the "Directeur d\'{e}partemental de la protection des populations" under Permit Number: A690290402. This article has been written to comply with the ARRIVE guidelines for reporting animal research.

\subsection{Subjects and Data}
Two macaque monkeys (macaca mulatta, Monkey R, 17-year-old female left-handed weighing 7 kg, and Monkey S, 16-year-old male right-handed weighing 8.5 kg) served as subjects. Prior to the experimental recordings, the monkeys were trained to perform the experimental task and then implanted for ECoG recordings. The data presented here have previously been subject to initial analysis in \cite{stoll_effects_2016} and as baseline data in \cite{wilson_prefrontal_2016}. As a result, the data are available at Open Science Framework, via the permanent link \url{osf.io/sdw2v}. The readme file describes the contents, and interested parties are welcome to contact CREW with questions about the data. Materials and methods for the implants and recordings are the same as described in those studies. We reiterate the key methodological components here. In the present analysis, we use 35 and 36 experimental sessions (for Monkey R and monkey S, respectively) that were collected on separate days. Monkeys occasionally take spontaneous breaks from work, as described in \cite{stoll_effects_2016}. Here, we only consider complete trials before the first break of the session, resulting in an average of 247 trials per session (minimum number of trials = 125, maximum of trials = 305) in monkey R and an average of 133 trials per session (minimum number of trials = 81, maximum of trials = 192) in monkey S.

\subsection{Electrode implants and choice of electrodes}
\begin{figure}[htbp]
\centering
\includegraphics[width = \linewidth]{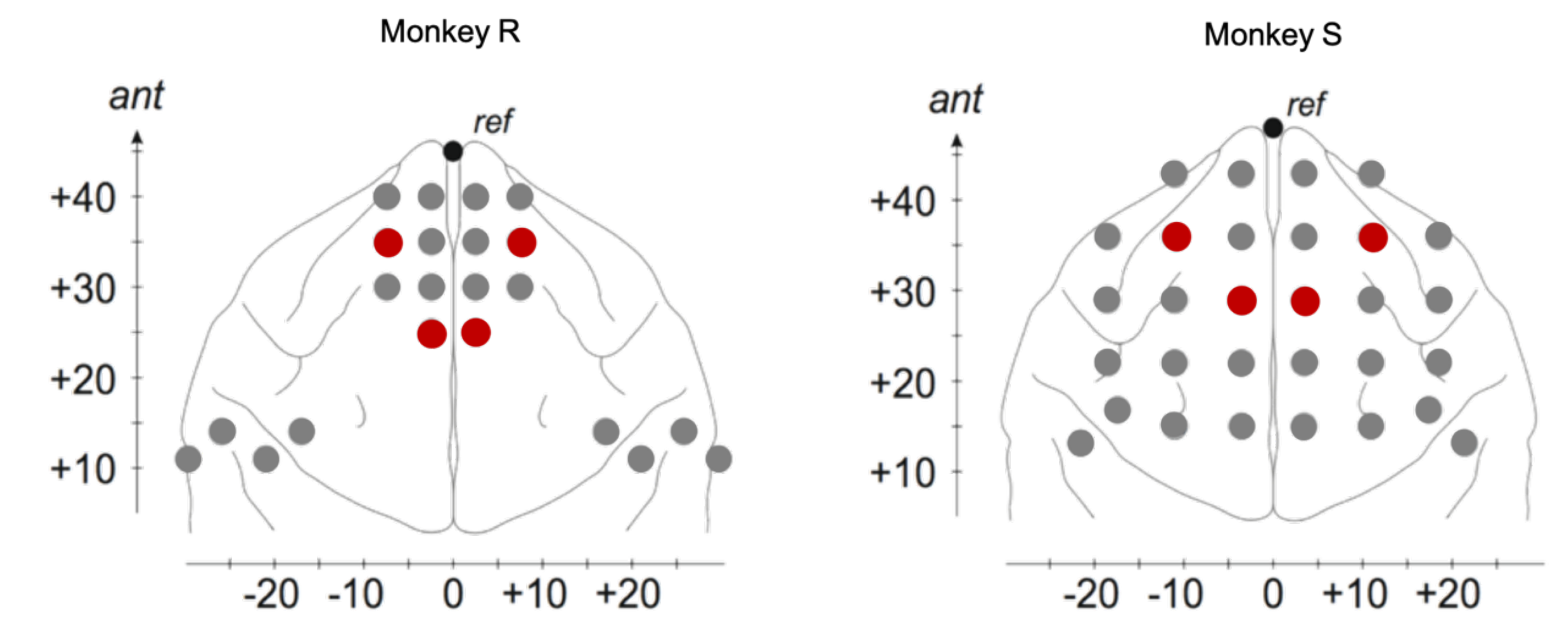}
\caption{\textit{Schematics of electrode grids.} The grey dots represent the location of the implanted electrodes. The black dot shows the reference electrode. Red dots represent electrodes chosen for the DCM analysis.}
\label{fig:electrodes}
\end{figure}

The implants comprised a head-post and a grid of transcranial electrodes for ECoG recordings, that is electrodes screwed through the skull to rest on the surface of the dura mater. More information about the surgical procedure can be found in \cite{stoll_effects_2016,wilson_prefrontal_2016}. Individual structural MRI images were used to select the implantation location of electrodes To ensure accuracy of implantation relative to anatomical landmarks, we used stereotaxic guidance to correctly place the electrode grids. Monkey R was implanted with a 5-mm-spaced grid of electrodes over the prefrontal cortex. In a second operation, eight electrodes over the sensorimotor cortex were added. Monkey S obtained a 7-mm-spaced grid of 31 electrodes in a single operation. The electrodes covered the prefrontal and sensorimotor cortex. For each monkey, the reference electrode was implanted into the think bone of the brow. An illustration of the electrodes can be seen in Figure \ref{fig:electrodes}. The electrodes representing the dlPFC are recorded within the Brodmann 9/46 areas \cite{petrides_dorsolateral_1999}, while the electrodes recording activity from the pre-SMA were placed in Brodmann area 6 \cite{luppino_corticocortical_1993,petrides_dorsolateral_1999}. The placement of the electrodes on the cortex, as well as the electrodes chosen for the analyses reported in this paper, are reported in figure \ref{fig:electrodes}.
A supplementary electrode which was screwed into the bone of the thick brow of the monkey on the midline anterior to the frontal grid served as reference.

\subsection{Experimental Design and Procedure}
\begin{figure}[htbp]
\centering
\includegraphics[width = \linewidth]{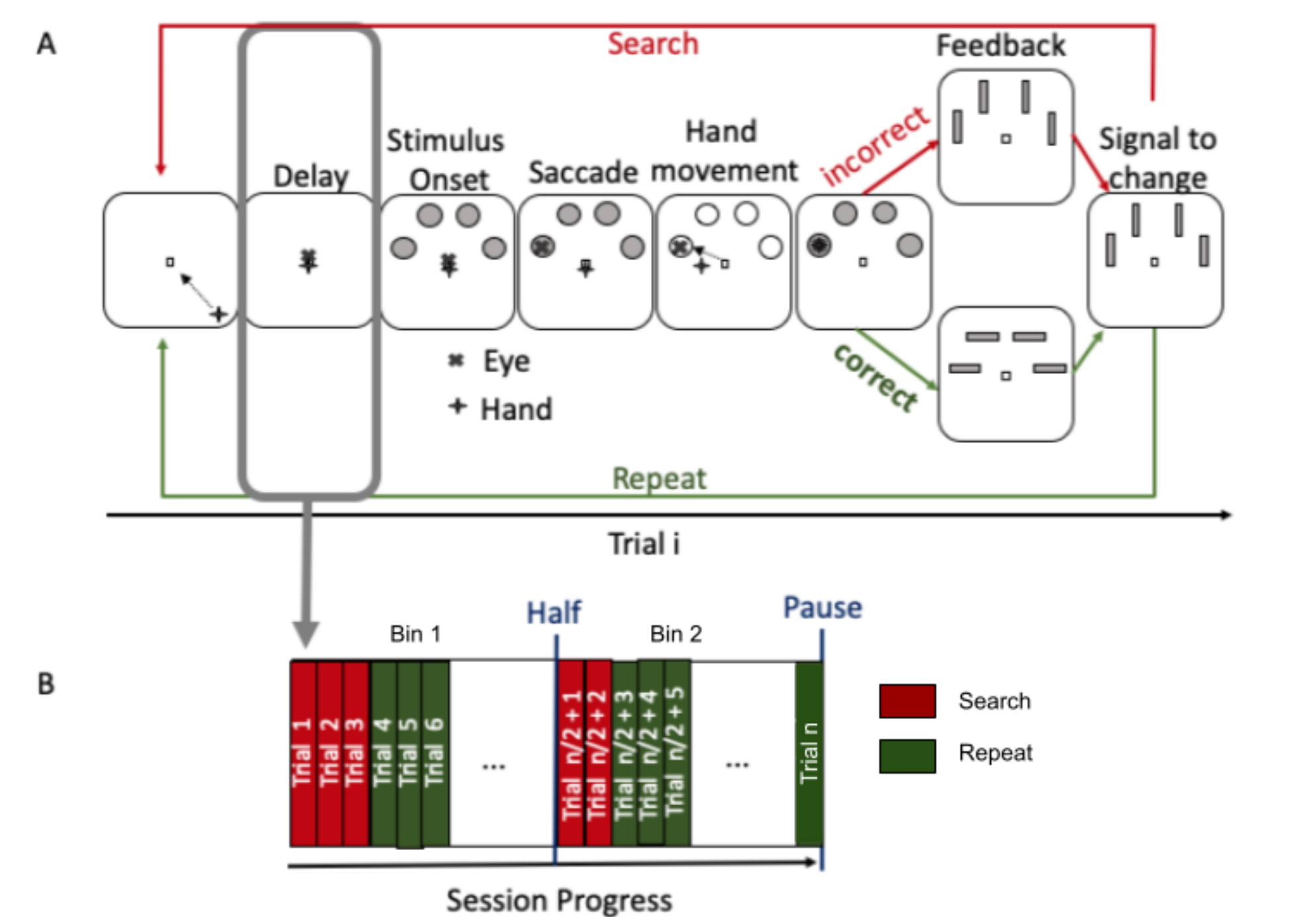}
\caption{\textit{A. Sequence of an individual trial.} The trial-and-error search among four stimuli and the selection of the identified target was rewarded and repeated up to three times. Frames show the consecutive events in each trial during a problem.\textit{B. Trial division during sessions.} For the frequency and connectivity analysis, we analyzed the ECoG recording during the delay period (marked by the grey frame in A) of each trial within a session until the first self-initiated break. Trials were divided into the first (bin 1) and the second half (bin 2) of the session according to their chronological order.}
\label{fig:design}
\end{figure}

Monkeys were seated in a primate chair in front of a touch-screen monitor and performed the problem-solving task. This task consists in a succession of ‘problems’ each including two phases, search and repeat, that differ in the level of cognitive control that they require \cite{procyk_modulation_2006}. A problem refers to a block in which the animal had to find—by trial and error, during the search phase—the rewarded target across several trials from four identical stimuli that were located in a semicircle above a fixation cross. Monkeys were rewarded with a 1-1.8 ml drop of juice upon selection of the correct target. After selecting a target, monkeys received visual feedback, and if the feedback was correct, a subsequent juice reward. When an incorrect target was chosen, monkeys received negative visual feedback and continued with the search phase on the next trial. The repeat phase commenced once the correct target had been discovered. Monkeys could repeat the selection of the rewarded target up to three times and be rewarded for those correct choices. Completion of the repeat phase ended the problem: the rewarded target was then re-randomized, a change signaled by a visual “signal to change”, and the next problem began with a new search phase.
Subjects initiated each trial by touching and holding a square on the screen at the bottom of the screen. A fixation point appeared, and fixating the fixation point while continuing to touch the square initiated the delay period of 1400 ms. Then four grey circular stimuli appeared in a semicircle above the fixation point. At the onset of the stimuli, monkeys had to saccade to the stimulus of their choice and fixate it during a period that was randomly varied among 400, 600, and 800 ms. The end of the fixation period was marked by a change of luminance of the stimuli from grey to white, signaling the subjects to confirm their selection with a hand movement to the stimulus. After a period lasting between 600 and 1200 ms (steps of 200 ms), subjects received visual feedback for 800 ms. Four horizontal or vertical rectangles located in the same position as the stimuli indicated correct and incorrect stimulus selection, respectively. After the completion of the search and consecutive repeat phase, the four vertical rectangles were flashed three times to indicate the start of a new problem. See figure \ref{fig:design}A for an illustration of the trial events.
The delay period (i.e., the time period preceding the stimulus onset in the problem-solving task), is thought to include the preparation of an upcoming decision for a stimulus, during which information about previous trials is considered. As the search condition requires free decisions for a stimulus that demands more information than a mere repetition of the identified target, cognitive demands are thought to be higher for search trials than for repeat trials \cite{procyk_modulation_2006,khamassi_behavioral_2015}. Likewise, forming a decision can become more challenging with session duration, due to the need to compensate for fatigue. Since the delay varies with cognitive demands and increasing frontal activity is associated with task demands and time on task \cite{stoll_effects_2016,wilson_prefrontal_2016}, we used the delay period to estimate how effective connectivity varies as a function of task demands and time on task.

\subsection{Data Processing}
The signal was amplified and filtered (1–250 Hz) and digitized at 781.25 Hz. Data analysis was performed with homemade Matlab scripts using MATLAB 2017b and spm12. An initial stage of the data processing involved the decomposition of the ECoG signal with an independent component analysis, using the logistic infomax algorithm \cite{bell_information-maximization_1995}, in order to remove signal elements common to several areas of cortex, for example movement artefacts, or conduction by bone and dura mater. To this extent the signal is largely unique to each area, although the nature of the ECoG signal, recorded from the surface of the brain, precludes the complete exclusion of some common elements. Specifically for the two signals used in this analysis, the cross-correlation was always smaller than 0.15; in DCM, a small degree of common signal between nodes such as that caused by volume conduction is addressed by the fitting of the observation model.  To quantify changes in behavior, power, and effective connections across task demands and time on task, we distinguished between levels of task demands (repeat vs. search) and levels of time on task (first bin vs. second bin) and used a 2 x 2 factorial design. For the levels of time on task, we divided trials into two halves according to their chronological order (see figure \ref{fig:design}B). In the case of an odd number of trials, the center trial was assigned to the second half of the trials. 

\subsection{Behavioral Analysis}
\subsubsection{Task Engagement} Since cognitive effort is defined as the activity that compensates for the costs of task engagement, we needed to establish that monkeys engaged in the task throughout the experiment. We consider the percentage of rewarded stimulus selection as an indicator of the level of task engagement. Selecting an incorrect stimulus twice during search periods was considered as nonoptimal because the reward association is deterministic. In order to ensure that cognitive performance remained stable throughout the session, we compared the rate of nonoptimal choices in complete search trials in the first to the second half of the session using a right-sided Wilcoxon signed rank test. A significant increase would violate our assumption that monkeys remained engaged in the task. As the monkeys were trained prior to the experiment, we expect an overall low level (~10\%) of nonoptimal stimulus trials.
\subsubsection{Vigilance Decrement}. The continuous task execution is thought to be accompanied by progressive mental fatigue. Reaction time was measured as the time between the stimulus onset and the touch of the selected stimulus and its increase over the session is assumed to be indicative of increasing fatigue. To test whether reaction time increases over time and, implicitly, whether monkeys experienced mental fatigue as the session progressed, we performed a two-way ANOVA for reaction time using task demands and time on task as independent variables.

\subsection{DCM Analysis}
\begin{figure}[htbp]
\centering
\includegraphics[width = \linewidth]{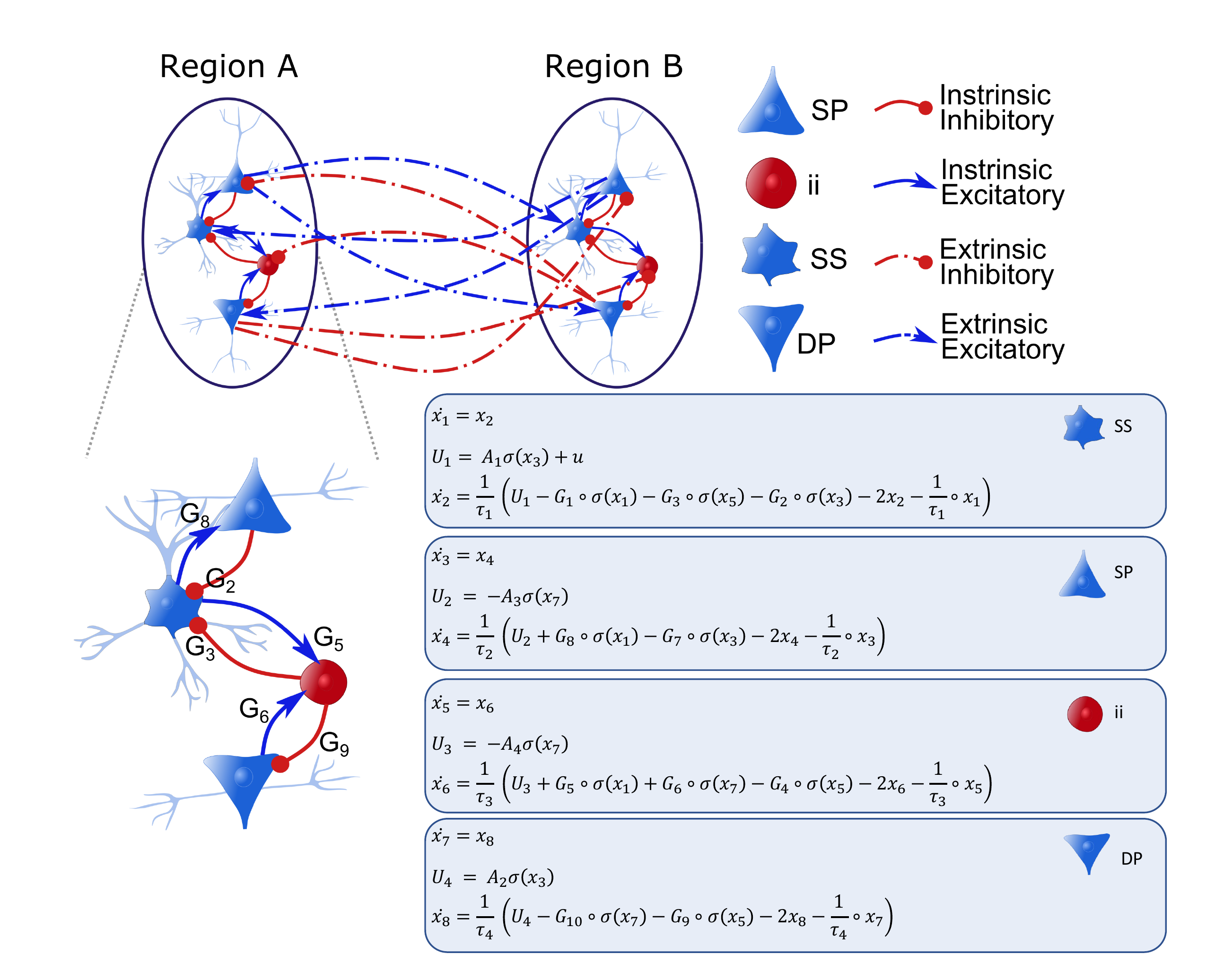}
\caption{\textit{Illustration of the canonical microcircuit.} The figure shows a connectivity model with connections from region A to region B. Forward connections (FW) originate from superficial pyramidal cells and target (1) spiny stellates and (2) deep pyramidal cells. Backward connections (BW) originate from deep pyramidal cells and target (1) superficial pyramidal cells and (2) inhibitory interneurons. In the models used here, there are FW and BW connection from A to B and from B to A. In the equations , FW excitatory connections are A\textsubscript{1} and A\textsubscript{2} and BW inhibitory connections are A\textsubscript{3} and A\textsubscript{4}. In addition, the figure displays the schematic of the intrinsic connections (i.e. the connectivity between the different populations within a source) The equations describe the dynamics of the different subpopulation. Here $\sigma$ is a sigmoidal activation function which transform post-synaptic depolarization into pre-synaptic firing rate and $\tau$ is the synaptic time constant.}
\label{fig:model}
\end{figure}

Frequency Analysis. In order to estimate cross-spectral densities, we used Bayesian multivariate autoregressive modeling \cite{penny2002bayesian} The peculiarity of this Bayesian MVAR modelling is the search for the optimal model order allowing a tradeoff between accuracy and complexity. Furthermore, the number of hidden states per source places an upper bound on the order of the VAR model. An extended discussion on these aspects is found in \cite{moran_dynamic_2009}. Following the arguments presented therein, an order of p=8 was chosen for our analyses. Using a frequency interval between 3 to 40 Hz, we estimated cross-spectral power in the dlPFC and pre-SMA for each monkey, hemisphere, session, and level of task demands and time-on-task.
The DCM framework includes a check on the explained variance, raising a flag every time that the model appears not to fit well despite convergence. No flags were raised in these analyses. The lags in the MVAR model determine the smoothness of the spectra. Again, there is an optimal space between accuracy and precision, and avoidance of overfitting. (Cross-)spectral densities were normalized between sessions by dividing by the maximum value.
It is worth to note that in DCM the shape of the observed spectra is determined by the parametrized 1/f neural fluctuations (a.k.a., innovations) and importantly the transfer functions that govern ‘spectral bumps’ in the output \cite{moran_dynamic_2009,friston_dcm_2012}.
We tested for power modulations with task demands and time on task—in the selected electrodes by means of a nonparametric test for dependent samples, the Wilcoxon signed rank test. The test was conducted for each monkey, hemisphere, and experimental variable (task demands and time on task), separately.
Model Inversion. For the estimation of effective connections between the dlPFC and pre-SMA underlying the changes reported above, we applied DCM for cross-spectral density based on the canonical microcircuit model (CMC; \cite{pinotsis_dynamic_2013,pinotsis_contrast_2014}. Effective connections are directed causal influences of one neural population on another \cite{friston_dynamic_2003}. DCM uses a Bayesian model inversion with cytoarchitecturally inspired neural mass models that are fitted to empirical observations (\cite{friston_dcm_2012,moran_neural_2013}). This allows one to estimate and make inferences about how neural subpopulations influence one another (\cite{kiebel_dynamic_2008}) using Bayesian model comparison. A model in this setting is a set of differential equations that describe the dynamics between coupled sources. Allowable connections are specified as a multivariate (log) normal density, in terms of means and variances. The prior distribution determines which connections between the sources are thought to exist a priori, and are thus fitted to the data to produce a posterior density. The model inversion procedure (known as Variational Laplace) estimates the posterior density of the model parameters in a way that furnishes the best trade-off between the accuracy and complexity. In other words, in maximising a lower (variational free energy) bound on marginal likelihood or model evidence one is implicitly finding an accurate account of the data that has the smallest complexity (i.e., the degrees of freedom used to explain the data).
Each source is modeled as a canonical microcircuit, in which the data are thought to be generated by four neural subpopulations that inhabit distinct cortical layers (\cite{moran_neural_2013}). These subpopulations are deep (dp) and superficial pyramidal (sp) cells, spiny stellate (ss) cells and inhibitory interneurons (ii). For between-source connections, the model distinguishes between forward and backward connections (\cite{felleman_distributed_1991}). Both kinds of connections differ in their origin and target subpopulations: Forward connections originate in superficial pyramidal neurons of one source and target the spiny stellate cells and deep pyramidal cells of another source. Backward connections originate in the deep pyramidal cells of one source and target the inhibitory interneurons and superficial pyramidal cells of another source. Figure \ref{fig:model}, shows the CMC model and how it is used here.There are forward (FW) and backward(BW) connections from A to B and vice versa. In addition, the connectivity schematic between the different populations within a source is shown together with the neural state equations of the different populations. Note that the model comprises four extrinsic connections in each direction, for more details please see \cite{pinotsis_working_2019}. For each monkey, session and condition, we inverted a model that was comprised eight extrinsic connections between dlPFC and pre-SMA. The (within condition) posterior estimates of each connection were then used for the analysis of (between condition) task demands and time effects. It is worth noting that the noise is estimated in the Variational Bayes scheme, handling differences in noise levels across sessions.

\subsection{Parameter Inference} 
\begin{figure}[htbp]
\centering
\includegraphics[width = \linewidth]{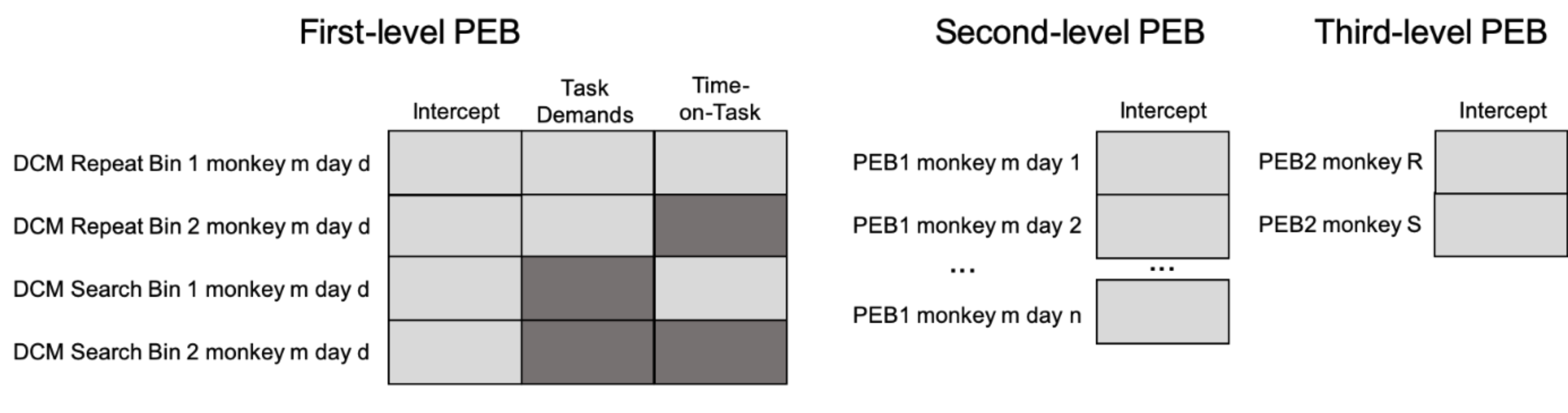}
\caption{\textit{PEB analysis.} In the first level, changes in effective connections across task and time levels are inferred for each session and subject. In the second level, the consistency across days for each subject is tested. At the third level, the consistency across subjects is evaluated. The third level PEB yields posterior probabilities that indicate the confidence that the respective parameter is consistently modulated by task demands and time.}
\label{fig:PEB}
\end{figure}

In order to identify effective connections between the prefrontal and premotor areas that are consistently modulated by task demands and time on task, we used a (between condition) hierarchical model within the parametric empirical Bayesian (PEB) framework \cite{friston_bayesian_2016,pinotsis_intersubject_2016,zeidman_guide_2019}. PEB uses a Bayesian general linear model, which we used to compare connectivity between experimental conditions (e.g., task demands) and to investigate the consistency of effect sizes (e.g., consistency of the demand effects across sessions). Since we were interested in (1) which connections were modulated by task demands and time on task, (2) which modulation effects are consistent across sessions, and (3) which modulation effects are consistent between monkeys, we conducted a hierarchical PEB analysis based with three levels: a hierarchical approach used in previous studies \cite{park_hierarchical_2017,van_de_steen_dynamic_2019}. The design matrices for the three levels are represented in figure \ref{fig:PEB}. 
The model fitting at the first level is reminiscent of the classical summary statistic approach to random effects, in which connection strengths are the dependent variables, and the experimental conditions are the independent variables. The PEB model, however, brings with it the advantage of including uncertainty about parameter estimates, thereby weighting the influence of the first level (condition specific) estimates on inferences at the second (between condition) levels. At the second level, for each monkey we tested whether the (between condition, within subject) effect sizes of task demands and time on task were consistent across session, while at the third level, we tested for (between subject) consistency between monkeys. As we are interested in only those connections that were consistently modulated across sessions and subjects, we assessed the evidence for consistent effects on connectivity at the third PEB level: we considered a posterior probability larger than 95\% to signify a relevant contribution of the given predictor. In other words, we compared models with and without an effect on a particular connection and used the ensuing estimates of model evidence (a.k.a., marginal likelihood) based upon variational free energy to ensure models with a change in a particular connection were 20 times more likely than models without a change.

\section{Results}
\subsection{Behavior}
\subsubsection{Task Engagement} The median percentage of nonoptimal choices was around 10\% for both subjects (Monkey R: Median = 10.19\%, Interquartile Range = 0.09\%; monkey S: Median = 10.00\%, Interquartile Range = 0.12\%).  In order to test for normality of the error rates, we performed a one-sample Shapiro-Wilk test. As we found a violation of normality for one subject (Monkey R: W = 0.972, p = 0.120; monkey S; W = 0.947, p = 0.004), we used the Wilcoxon signed rank test to investigate whether error rates during the search phase increased over time. Increases in error rates weren't significant (Monkey R: z = -0.05, p = 0.961; monkey S: z = 1.556, p = 0.12). There is, therefore, no evidence that task engagement decreased across time.
\subsubsection{Vigilance Decrement} Trials with reaction times larger than three standard deviations from the mean were considered outliers and removed from the analysis. The Shapiro-Wilk test was used to licence the use of an ANOVA. No violation for the normality assumption was detected (monkey R: W = 0.987, p = 0.219; monkey S: W = 0.985, p = 0.118). On average, the time between stimulus onset and the touch of the chosen stimulus was 700.95 ms (SD = 31.65) and 686.00 ms (SD = 37.00) for monkey R and S, respectively. Reaction time significantly increased with time on task (monkey R: F1,133 = 30.65, p < 0.001; monkey S: F1,141 = 19.5, p < 0.001). For monkey S, the reaction time was also longer for search trials than for repeat trials (monkey R: F1,133 = 0.16, p = 0.688; monkey S: F1,141 = 7.71, p = 0.006). Considering that reaction time increases with time on task, vigilance is, therefore, assumed to decline across sessions. 

\subsection{DCM Analysis}
\subsubsection{Frequency Analysis} 
\begin{figure}[htbp]
\centering
\includegraphics[width = \linewidth]{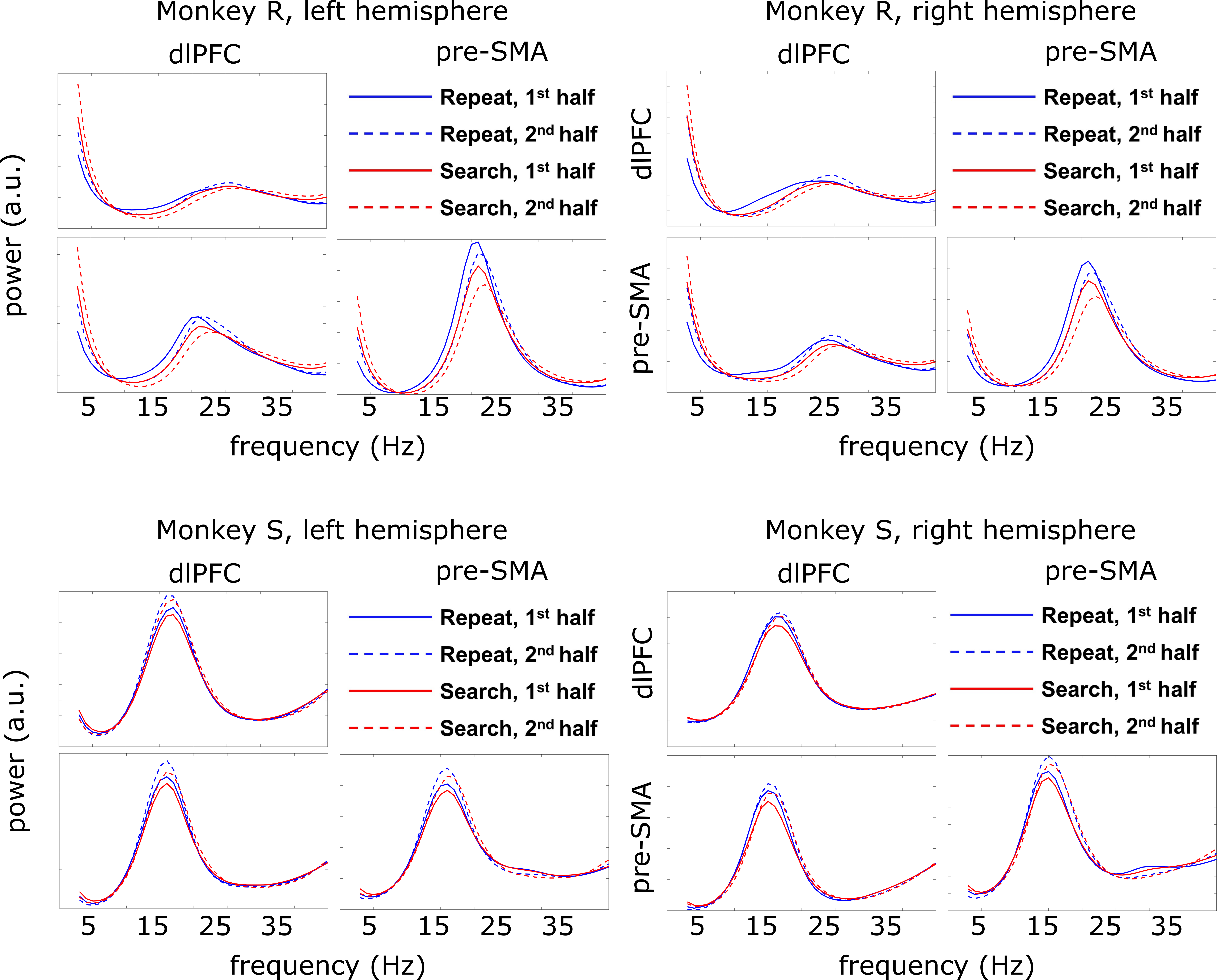}
\caption{\textit{Channels auto- and cross-spectral density}. Auto- and cross-spectral density for each task condition for each channel, monkey, and hemisphere, averaged over the first and second half of the trials. A linear polinomial fit of order 1 was removed from the natural logarithm of the spectra, before converting back to linear scale, for visualization purposes. Differences are quantified in Table \ref{tab:broadband}.}
\label{fig:spectra}
\end{figure}

The auto- and cross-spectral densities for the two monkeys, hemispheres, regions, and conditions, averaged over two separate portions of the trials (first and second half), are displayed in figure \ref{fig:spectra}. Before testing whether power increased with task demands and time on task between the two portions of the trials, we performed a Shapiro-Wilk test for normality. Across all (cross-)spectra and monkeys we performed eight tests and adjusted the significance level accordingly (Bonferroni corrected alpha = 0.0063). Because we found a violation of normality for some (cross-)spectra (see Table \ref{tab:broadband}), we used the Wilcoxon signed rank test to assess the increase in power between task demands and time on task. We further asked whether power increased in search versus repeat trials and in the second versus the first half of the experimental session before the first pause. The positive test statistics throughout all electrode sites, monkeys and experimental variables—as displayed in Table \ref{tab:broadband}—indicate an increase with task demands and time on task. This time-related increase in power was significant for all (cross-)spectra and monkeys. The task-related increase in power was significant for monkey R only, and only for dlPFC, in both hemispheres Tables \ref{tab:alphaband} and \ref{tab:betaband} in the Appendix report the same tests in specific bands. 

\begin{table}[ht!]
\begin{tabular}{clccc
>{\columncolor[HTML]{FFFFFF}}c ccc
>{\columncolor[HTML]{FFFFFF}}c cc}
                          &                                  & \multicolumn{2}{c}{\cellcolor[HTML]{F9CB9C}Normality} & \multicolumn{8}{c}{\cellcolor[HTML]{A6E3B7}Broadband power   increase}                                                                                                                                                                                     \\
                          &                                  & \multicolumn{2}{c}{\cellcolor[HTML]{F9CB9C}}          & \multicolumn{4}{c}{\cellcolor[HTML]{A6E3B7}Task}                                                                                        & \multicolumn{4}{c}{\cellcolor[HTML]{A6E3B7}Time}                                                                 \\
                          &                                  & \cellcolor[HTML]{FCE5CD}z & \cellcolor[HTML]{FCE5CD}p & \cellcolor[HTML]{D1F1DA}MPC & \cellcolor[HTML]{D1F1DA}SE & \cellcolor[HTML]{D1F1DA}z & \cellcolor[HTML]{D1F1DA}p                        & \cellcolor[HTML]{D1F1DA}MPC & \cellcolor[HTML]{D1F1DA}SE & \cellcolor[HTML]{D1F1DA}z & \cellcolor[HTML]{D1F1DA}p \\
\cellcolor[HTML]{F2F2F2}R & \cellcolor[HTML]{F2F2F2}L dlPFC  & \textbf{0.86}             & \textbf{\textless{}1E-3}  & \textbf{-2.56\%}            & \textbf{0.007}             & \textbf{-3.161}           & \cellcolor[HTML]{FFFFFF}\textbf{\textless{}1E-3} & \textbf{8.47\%}             & \textbf{0.011}             & \textbf{4.23}             & \textbf{\textless{}1E-3}  \\
\cellcolor[HTML]{F2F2F2}  & \cellcolor[HTML]{F2F2F2}L preSMA & \textbf{0.87}             & \textbf{\textless{}1E-3}  & 0.75\%                      & 0.005                      & 1.245                     & \cellcolor[HTML]{FFFFFF}0.213                    & \textbf{6.96\%}             & \textbf{0.007}             & \textbf{4.68}             & \textbf{\textless{}1E-3}  \\
\cellcolor[HTML]{F2F2F2}  & \cellcolor[HTML]{F2F2F2}L cross  & \textbf{0.85}             & \textbf{\textless{}1E-3}  & -0.73\%                     & 0.007                      & -0.835                    & \cellcolor[HTML]{FFFFFF}0.404                    & \textbf{9.48\%}             & \textbf{0.011}             & \textbf{4.21}             & \textbf{\textless{}1E-3}  \\
\cellcolor[HTML]{F2F2F2}  & \cellcolor[HTML]{F2F2F2}R dlPFC  & \textbf{0.85}             & \textbf{\textless{}1E-3}  & \textbf{-3.41\%}            & \textbf{0.009}             & \textbf{-3.669}           & \textbf{\textless{}1E-3}                         & \textbf{8.82\%}             & \textbf{0.012}             & \textbf{4.47}             & \textbf{\textless{}1E-3}  \\
\cellcolor[HTML]{F2F2F2}  & \cellcolor[HTML]{F2F2F2}R preSMA & \textbf{0.85}             & \textbf{\textless{}1E-3}  & 0.83\%                      & 0.005                      & 1.441                     & \cellcolor[HTML]{FFFFFF}0.150                    & \textbf{6.46\%}             & \textbf{0.007}             & \textbf{4.62}             & \textbf{\textless{}1E-3}  \\
\cellcolor[HTML]{F2F2F2}  & \cellcolor[HTML]{F2F2F2}R cross  & \textbf{0.85}             & \textbf{\textless{}1E-3}  & -2.54\%                     & 0.011                      & -1.589                    & \cellcolor[HTML]{FFFFFF}0.112                    & \textbf{11.00\%}            & \textbf{0.016}             & \textbf{4.23}             & \textbf{\textless{}1E-3}  \\
\cellcolor[HTML]{F2F2F2}S & \cellcolor[HTML]{F2F2F2}L dlPFC  & 0.99                      & 0.205                     & 1.72\%                      & 0.011                      & 0.992                     & \cellcolor[HTML]{FFFFFF}0.321                    & \textbf{9.02\%}             & \textbf{0.011}             & \textbf{4.66}             & \textbf{\textless{}1E-3}  \\
\cellcolor[HTML]{F2F2F2}  & \cellcolor[HTML]{F2F2F2}L preSMA & 0.98                      & 0.043                     & 1.75\%                      & 0.011                      & 1.224                     & \cellcolor[HTML]{FFFFFF}0.221                    & \textbf{8.82\%}             & \textbf{0.011}             & \textbf{4.70}             & \textbf{\textless{}1E-3}  \\
\cellcolor[HTML]{F2F2F2}  & \cellcolor[HTML]{F2F2F2}L cross  & 0.98                      & 0.121                     & 1.71\%                      & 0.012                      & 0.938                     & \cellcolor[HTML]{FFFFFF}0.348                    & \textbf{9.31\%}             & \textbf{0.012}             & \textbf{4.64}             & \textbf{\textless{}1E-3}  \\
\cellcolor[HTML]{F2F2F2}  & \cellcolor[HTML]{F2F2F2}R dlPFC  & 0.97                      & 0.027                     & 0.66\%                      & 0.005                      & 1.153                     & \cellcolor[HTML]{FFFFFF}0.249                    & \textbf{3.08\%}             & \textbf{0.005}             & \textbf{3.82}             & \textbf{\textless{}1E-3}  \\
\cellcolor[HTML]{F2F2F2}  & \cellcolor[HTML]{F2F2F2}R preSMA & 0.98                      & 0.294                     & 0.51\%                      & 0.007                      & 0.721                     & \cellcolor[HTML]{FFFFFF}0.471                    & \textbf{4.00\%}             & \textbf{0.006}             & \textbf{4.04}             & \textbf{\textless{}1E-3}  \\
\cellcolor[HTML]{F2F2F2}  & \cellcolor[HTML]{F2F2F2}R cross  & \textbf{0.96}             & \textbf{0.004}            & -0.90\%                     & 0.007                      & -0.601                    & \cellcolor[HTML]{FFFFFF}0.548                    & \textbf{3.30\%}             & \textbf{0.007}             & \textbf{3.48}             & \textbf{\textless{}1E-3} 
\end{tabular}
\caption{Tests for broadband power for monkeys R and S, for spectra and cross-spectra in the left (L) and right (R) hemisphere. MPC = Mean Percentage Change; SE = Standard Error. Values in bold indicate significance at a Bonferroni-corrected level of 0.0063}\label{tab:broadband}
\end{table}

\subsection{Changes in effective connectivity } 
\begin{figure}[htbp]
\centering
\includegraphics[width = \linewidth]{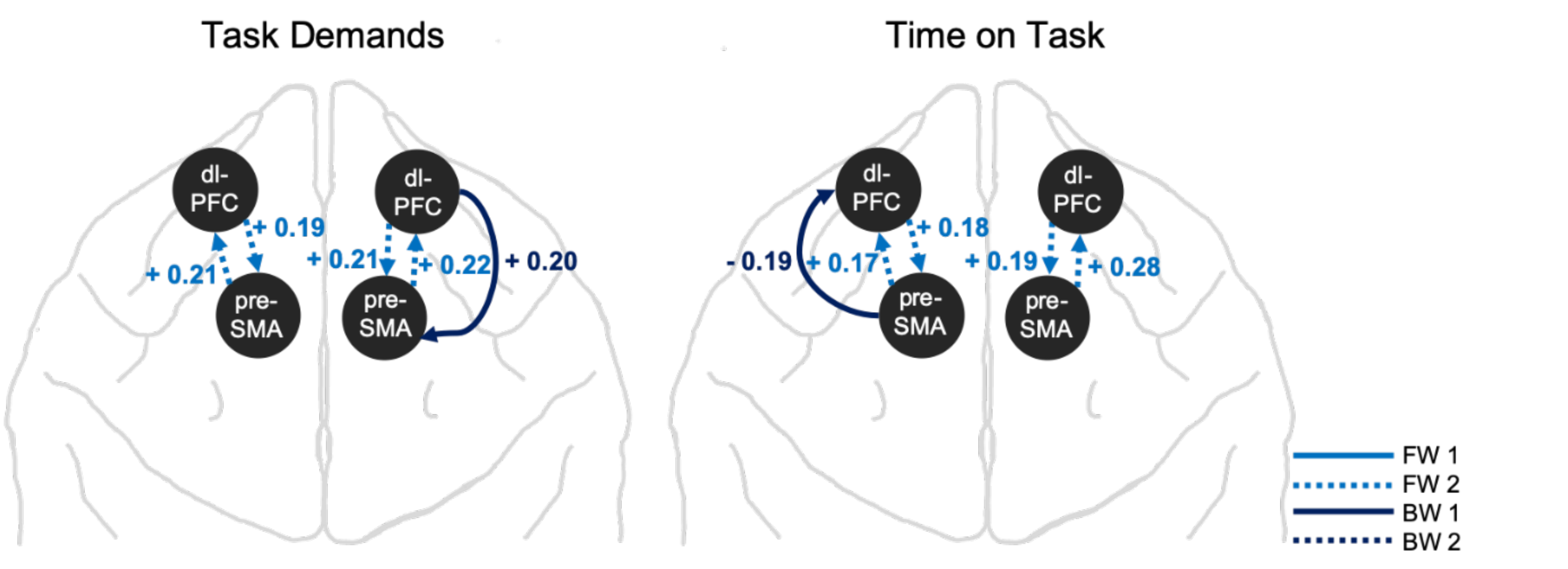}
\caption{\textit{Effective connectivity modulations.} Depicted connections (as log scaling factors) between the dlPFC and pre-SMA with changes that have a posterior probability larger than 95\%. Left panel shows modulations related to task demands, while the right panel shows time-induced connectivity modulations. Light blue arrows represent forward connections and dark blue arrows depict backward connections. Each connection is accompanied by its effect size (i.e., change in connectivity strength related to the experimental variable). FW1: Connections from superficial pyramidal  cells targeting spiny stellate cells; FW2: targeting  deep pyramidal cells. BW1: Connections from deep pyramidal cells targeting superficial pyramidal cells; BW2:  inhibitory interneurons. See also Figure \ref{fig:model}.}
\label{fig:DCMres}
\end{figure}

The aim of the analysis was to test whether the increase in task demands and time on task modulate effective connectivity between the dlPFC and pre-SMA and whether these two aspects of cognitive effort manifest in the same effective connections. We were interested in whether both areas (dlPFC, pre-SMA) and hemispheres (left and/or right) would evince changes in effective connectivity, and whether these changes would be evident in feedforward, feedback or lateral connections between brain areas. We compared all possible models assessing connectivity changes as a result of task demands and time on task . Using Parametric Empirical Bayes (PEB) \cite{friston_dcm_2014,pinotsis_intersubject_2016,zeidman_guide_2019} we assess the evidence for condition specific changes in all connectivity parameters that were conserved over subjects. Figure \ref{fig:DCMres} illustrates the effective connections that were modulated by task demands (left panel) or time on task (right panel) with a posterior probability of greater than 95\%. 
Both condition specific effects change effective connectivity. An increase in task demands is accompanied by an increase in forward connections targeting deep pyramidal cells (FW2) and an increase in the backward connection from the right dlPFC to the right pre-SMA targeting superficial pyramidal cells (BW1). Forward connections in both directions, anterior-to-posterior and posterior-to-anterior, are affected. Time on task is, likewise, related to an increase in forward connections (FW2) in both directions and, in addition, to a decrease in the backward connection from the left pre-SMA to the spiny stellate cells of the left dlPFC (BW1). Both experimental effects (demands and time) have overlapping effects in terms of the type and direction of the connection and the direction of the modulatory effect: both task demands and time on task affect forward connections in both directions.  Effective connections increase in strength when task demands and time on task increase. While both experimental variables modulate the same type of backward connection, the hemisphere, and direction of the connection and of the modulation effect are distinct. These differences could be explained by the different roles of dlPFC and pre-SMA in processing cognitive load and time as discussed below.

\section{Discussion} 
The study provides novel insights into the shared changes in connectivity induced by task demands and time on task within the FPN. Forward connections between the dlPFC and pre-SMA increased in strength when task demands were high, and as the experimental session progressed. The increase in connectivity as a consequence of task demands replicates previous computational \cite{edin_mechanism_2009,wei_distributed_2012} and empirical findings \cite{koechlin_architecture_2003,ma_working_2012,dima_dynamic_2014,harding_effective_2015,pinotsis_working_2019}. A similar increase of forward and backward connections from PFC with increasing load was found in \cite{pinotsis_working_2019} and was attributed to higher feedforward drive due to higher load and increasing stabilizing signals to counteract the increase in cognitive demands. This is also in agreement with the theory of Predictive Coding, which suggests that feedforward and feedback connections carry prediction errors and predictions of sensory information as two counterstreams of information; one upstream and one downstream \cite{friston_free-energy_2009,pinotsis_working_2019}. Downstream predictions from PFC might increase to balance increasing task demands and upstream prediction errors. The \textit{predictive} nature of the spectral responses analysed here are underlined by the fact that data are taken from the delay period before the onset of the trial. The behavioural encoding is therefore predictive of the content of the trial about to start - for more detail on this see \cite{stoll_effects_2016}.

Our contribution to the field is to show increasing fronto-frontal connectivity strength with time on task, and to compare time-on-task connectivity modulations with the effects of task demands. Interestingly, with time-on-task the feedback from pre-SMA to dlPFC decreased, which could indicate decreased temporal processing as a result of fatigue; pre-SMA is known to be implicated in perceptual timing, rhythm perception \cite{chen_listening_2008}, time networks \cite{meck_cortico-striatal_2008} and sensory temporal processing, see the metareview by Schwartze et al. \cite{schwartze_functional_2012}. This is also in agreement with the theory of Predictive Coding;  with increasing fatigue (longer time on task), timing prediction signals from pre-SMA  might attenuate; leading to impaired temporal processing and decreased rates of evidence accumulation. This impairment can also explain increased upstream prediction error signals:  we show that forward connections originating in both pre-SMA and dlPFC, increased with time on task. This is similar to the increase in forward connections when cognitive demands increased, found above. Although there is evidence for anatomical connections between the dlPFC and pre-SMA in both directions \cite{markov_weighted_2014}, previous studies have focused on unidirectional modulations of connections from the prefrontal to premotor areas. 

On a behavioral level, monkeys maintained good performance across time in terms of error rates but showed decreasing in reaction speeds. Two accounts may explain the deterioration on the one hand and the stability of performance over time on the other hand. First, resource theory (where \textit{resource} denotes the amount of effort that can be dedicated to a task) ascribes the deterioration of performance to mental fatigue \cite{ackerman_cognitive_2011,hopstaken_multifaceted_2015}. Mental fatigue increases when a cognitive task is executed over time because the pool of cognitive resources steadily depletes. In the present data, we observe signs of mental fatigue by means of a reaction time slowing, but not in cognitive errors. Why monkeys can maintain the accuracy of their performance in spite of emerging fatigue may be explained by motivational control theory. According to this account, the decision to engage in a cognitive task is the result of a cost-benefit analysis \cite{sarter_more_2006,hopstaken_shifts_2016,umemoto_electrophysiological_2019}. Task performance is only maintained if a cost-benefit analysis shows that the predicted rewards outweigh the costs \cite{hockey_compensatory_1997}. The cost is the effort needed in order to compensate the level of mental fatigue, while the benefit is the level of predicted gratification. In the current experiment, monkeys were rewarded with juice for every correct target selection and thus upholding the accuracy level led to a higher reward likelihood. In the face of emerging fatigue, monkeys thus increase their effort to maintain the rewarded aspect (i.e., accuracy) of the performance, while the unrewarded aspect (i.e., reaction speed) of performance deteriorated. This speaks to an accuracy speed trade-off that maintains reward likelihood.

These behavioral results suggest that two competing mechanisms, mental fatigue and cognitive effort, are in play. In the context of predictive coding, they are reconciled as different potential causes that lead to the same result: increased upstream prediction error signals. In other words, the increase in effective connectivity strength over time could  be attributed to mental fatigue or to its mitigation through an increase of effort, as mediated by changes in attentional set. In predictive coding, this would manifest as an increase in the precision or gain afforded prediction errors, thereby increasing their influence on Bayesian belief updating in the prefrontal cortex. One could argue that the overlap between the effects of task difficulty and time on task can be attributed, parsimoniously, to effortful changes in attentional set, rather than a mental fatigue effect. On this view, sending connections to the dlPFC from the pre-SMA may therefore increase in strength because the cognitive (i.e., attentional) effort—that is needed to maintain performance—increases with task difficulty and time on task. This interpretation in line with previous findings considered in terms of working memory \cite{edin_mechanism_2009,pinotsis_working_2019}.

Previous studies \cite{kornblith_stimulus_2016,stoll_effects_2016} have shown the association of frontal beta-band power and cognitive effort. Because beta-band power increased both with task demands and time on task, beta oscillations were thought to relate to changes in cognitive effort \cite{stoll_effects_2016}. This interpretation of the role of frontal beta-band oscillations is consistent with other studies, in which beta-band oscillations are ascribed to goal-directed cognition \cite{siegel_spectral_2012,antzoulatos_synchronous_2016,riddle_causal_2019}, and maintaining the task context in working memory \cite{engel_beta-band_2010}. Our study shows that the levels of cognitive effort are associated with forward connections between prefrontal and premotor areas. When considered in light of previous studies, this could constitute an indication that beta-band oscillations could reflect the enabling of forward connections. This fits comfortably with related treatments of beta-band suppression and rebound during movement, that has been interpreted in terms of sensory attenuation \cite{palmer_sensorimotor_2019}; namely, the attenuation of the precision of proprioceptive prediction errors during movement and the subsequent attention to the consequences of movement.

From studies of visual processing, we observe that visual information processing follows a functional hierarchy and that information from lower to higher-order regions may be mediated by high frequency rhythms, while transmission from higher to lower-order regions is associated with low frequency rhythms \cite{bosman_attentional_2012,kerkoerle_alpha_2014,pinotsis_contrast_2014,bastos_visual_2015}. Consistent with these connection asymmetries, Buffalo et al. \cite{buffalo_laminar_2011} found differences of low and high frequency oscillations within the cortex during visual processing. Some of these studies—that investigated the laminar-specific and asymmetries in oscillatory frequency—found beta-band oscillations to behave as such a low frequency rhythm \cite{buffalo_laminar_2011,bastos_visual_2015}. In these studies, beta-band oscillations were thus found in the deep layers of the cortex and drove backward connections. One might speculate that lower frequencies (such as alpha and beta) are responsible for mediating attentional set through—in the setting of predictive coding—predicting the precision of ascending prediction errors that are conveyed by forward connections.
As a first step in this direction, here we focused on a broad band rather than restricting ourselves to the beta one, considering that bands vary at an individual level (evidence also reflected in splitting beta in beta 1 and beta 2 in \cite{stoll_effects_2016}. 

\subsection{Limitations} A number of limitations of our study need to be mentioned and might provide directions for future work. First, cognitive effort (or attentional set) was not manipulated directly but rather interpreted as the result of the duration of task engagement and of task demands. It is important to discriminate between how demanding a task is and how much effort is deployed to overcome this challenge in order to distinguish between cognitive effort and mental fatigue effects (see \cite{agrawal2022temporal} for a recent review). We claim, in accordance with \cite{sarter_more_2006}, that only if the subject is motivated to perform the task, task demands and time on task may correlate with levels of cognitive effort. To measure motivation levels, future studies may include neural correlates of motivation or experimental manipulations of expected reward. Furthermore, the restriction to two cortical areas only provides a limited picture of the neural interactions that underwrite cognitive effort. The investigation of effective connectivity changes within an extended fronto-parietal network may afford better insight into effort effects in task-specific areas. In addition, although we showed robust effects across many days and two monkeys, future work needs to confirm the present results and generalize the effects to human subjects. 

In sum, this work highlights the common changes in connectivity associated with task demands and time on task effects within the FPN. The shared connectivity dynamics may underlie cognitive effort and deployment of attention that is required when the task becomes more difficult and when mental fatigue needs to be mitigated, to maintain performance. These insights contribute to the possible distinction between the neural mechanisms associated with cognitive effort versus mental fatigue and motivate future work on comparing the neural correlates of task demand and time on task effects.

\section{Acknowledgements}
KJF and DAP acknowledge financial support from UK Research and Innovation - UKRI ES/T01279X/1.
KW acknowledges support from the Research Foundation Flanders (FWO) and the Faculty of Psychology of Educational Sciences of Ghent University for mobility grants.
DM, EP, CREW, KW acknowledge support from Campus France and the Research Foundation Flanders (FWO) through the Tournesol exchange program.
FVDS and DM acknowledge support from Ghent University Research Council, grant number BOF17/GOA/004.
FVDS is funded by the Research Foundation – Flanders (FWO, project No. 1267422N).

\bibliography{Bibliography}
\newpage


\section*{Appendix}
\subsection*{Tables for alpha and beta power normality and changes}
\captionsetup{labelformat=AppendixTables}
\setcounter{table}{0}
\begin{table}[ht!]
\begin{tabular}{clccc
>{\columncolor[HTML]{FFFFFF}}c ccc
>{\columncolor[HTML]{FFFFFF}}c cc}
\multicolumn{1}{l}{}      &                                  & \multicolumn{2}{c}{\cellcolor[HTML]{F9CB9C}Normality} & \multicolumn{8}{c}{\cellcolor[HTML]{A6E3B7}Alpha power increase}                                                                                                                                                                                   \\
\multicolumn{1}{l}{}      &                                  & \multicolumn{2}{c}{\cellcolor[HTML]{F9CB9C}}          & \multicolumn{4}{c}{\cellcolor[HTML]{A6E3B7}Task}                                                                             & \multicolumn{4}{c}{\cellcolor[HTML]{A6E3B7}Time}                                                                    \\
\multicolumn{1}{l}{}      &                                  & \cellcolor[HTML]{FCE5CD}z & \cellcolor[HTML]{FCE5CD}p & \cellcolor[HTML]{D1F1DA}MPC             & \cellcolor[HTML]{D1F1DA}SE & \cellcolor[HTML]{D1F1DA}z & \cellcolor[HTML]{D1F1DA}p & \cellcolor[HTML]{D1F1DA}MPC & \cellcolor[HTML]{D1F1DA}SE & \cellcolor[HTML]{D1F1DA}z    & \cellcolor[HTML]{D1F1DA}p \\
\cellcolor[HTML]{F2F2F2}R & \cellcolor[HTML]{F2F2F2}L dlPFC  & 0.97                      & 0.014                     & -1.23\%                                 & 0.0082                     & -1.28                     & 0.200                     & \textbf{11.07\%}            & \textbf{0.012}             & \textbf{4.78}                & \textbf{\textless{}1E-3}  \\
\cellcolor[HTML]{F2F2F2}  & \cellcolor[HTML]{F2F2F2}L preSMA & 0.98                      & 0.166                     & \cellcolor[HTML]{FFFFFF}2.11\%          & 0.0069                     & 2.83                      & 0.004                     & \textbf{10.82\%}            & \textbf{0.008}             & \textbf{5.14}                & \textbf{\textless{}1E-3}  \\
\cellcolor[HTML]{F2F2F2}  & \cellcolor[HTML]{F2F2F2}L cross  & 0.98                      & 0.062                     & \cellcolor[HTML]{FFFFFF}1.63\%          & 0.0095                     & 1.91                      & 0.055                     & \textbf{13.23\%}            & \textbf{0.013}             & \textbf{5.00}                & \textbf{\textless{}1E-3}  \\
\cellcolor[HTML]{F2F2F2}  & \cellcolor[HTML]{F2F2F2}R dlPFC  & 0.99                      & 0.695                     & \cellcolor[HTML]{FFFFFF}-2.26\%         & 0.0085                     & -2.34                     & 0.019                     & \textbf{12.92\%}            & \textbf{0.012}             & \textbf{4.68}                & \textbf{\textless{}1E-3}  \\
\cellcolor[HTML]{F2F2F2}  & \cellcolor[HTML]{F2F2F2}R preSMA & 0.99                      & 0.328                     & \cellcolor[HTML]{FFFFFF}1.55\%          & 0.0068                     & 2.56                      & 0.011                     & \textbf{10.20\%}            & \textbf{0.008}             & \textbf{5.11}                & \textbf{\textless{}1E-3}  \\
\cellcolor[HTML]{F2F2F2}  & \cellcolor[HTML]{F2F2F2}R cross  & 0.99                      & 0.366                     & \cellcolor[HTML]{FFFFFF}-2.81\%         & 0.0137                     & -1.43                     & 0.154                     & \textbf{14.98\%}            & \textbf{0.018}             & \textbf{4.65}                & \textbf{\textless{}1E-3}  \\
\cellcolor[HTML]{F2F2F2}S & \cellcolor[HTML]{F2F2F2}L dlPFC  & 0.97                      & 0.014                     & \cellcolor[HTML]{FFFFFF}\textbf{2.03\%} & \textbf{0.0069}            & \textbf{2.94}             & \textbf{0.003}            & \textbf{5.35\%}             & \textbf{0.007}             & \textbf{4.45}                & \textbf{\textless{}1E-3}  \\
\cellcolor[HTML]{F2F2F2}  & \cellcolor[HTML]{F2F2F2}L preSMA & 0.98                      & 0.166                     & \cellcolor[HTML]{FFFFFF}\textbf{2.41\%} & \textbf{0.0067}            & \textbf{3.28}             & \textbf{0.001}            & \textbf{5.00\%}             & \textbf{0.007}             & \textbf{4.45}                & \textbf{\textless{}1E-3}  \\
\cellcolor[HTML]{F2F2F2}  & \cellcolor[HTML]{F2F2F2}L cross  & 0.98                      & 0.062                     & \textbf{2.50\%}                         & \textbf{0.0071}            & \textbf{3.28}             & \textbf{0.001}            & \textbf{5.23\%}             & \textbf{0.007}             & \textbf{4.39}                & \textbf{\textless{}1E-3}  \\
\cellcolor[HTML]{F2F2F2}  & \cellcolor[HTML]{F2F2F2}R dlPFC  & 0.99                      & 0.695                     & 1.76\%                                  & 0.0062                     & 2.31                      & 0.021                     & \textbf{2.34\%}             & \textbf{0.005}             & \textbf{3.32}                & \textbf{\textless{}1E-3}  \\
\cellcolor[HTML]{F2F2F2}  & \cellcolor[HTML]{F2F2F2}R preSMA & 0.99                      & 0.328                     & 1.30\%                                  & 0.0066                     & 1.49                      & 0.136                     & 1.45\%                      & 0.006                      & 1.75                         & 0.080                     \\
\cellcolor[HTML]{F2F2F2}  & \cellcolor[HTML]{F2F2F2}R cross  & 0.99                      & 0.366                     & 1.72\%                                  & 0.0070                     & 1.95                      & 0.052                     & 1.82\%                      & 0.006                      & \cellcolor[HTML]{FFFFFF}2.31 & 0.021                    
\end{tabular}
\caption{Tests for alpha ($[8 - 12]$ Hz) power for monkeys R and S, for spectra and cross-spectra in the left (L) and right (R) hemisphere. MPC = Mean Percentage Change; SE = Standard Error. Values in bold indicate significance at a Bonferroni-corrected level of 0.0063}\label{tab:alphaband}
\end{table}

\begin{table}[ht!]
\begin{tabular}{clccc
>{\columncolor[HTML]{FFFFFF}}c ccc
>{\columncolor[HTML]{FFFFFF}}c cc}
\multicolumn{1}{l}{}      &                                  & \multicolumn{2}{c}{\cellcolor[HTML]{F9CB9C}Normality} & \multicolumn{8}{c}{\cellcolor[HTML]{A6E3B7}Beta power increase}                                                                                                                                                                                                    \\
\multicolumn{1}{l}{}      &                                  & \multicolumn{2}{c}{\cellcolor[HTML]{F9CB9C}}          & \multicolumn{4}{c}{\cellcolor[HTML]{A6E3B7}Task}                                                                                & \multicolumn{4}{c}{\cellcolor[HTML]{A6E3B7}Time}                                                                                 \\
\multicolumn{1}{l}{}      &                                  & \cellcolor[HTML]{FCE5CD}z & \cellcolor[HTML]{FCE5CD}p & \cellcolor[HTML]{D1F1DA}MPC             & \cellcolor[HTML]{D1F1DA}SE & \cellcolor[HTML]{D1F1DA}z & \cellcolor[HTML]{D1F1DA}p    & \cellcolor[HTML]{D1F1DA}MPC              & \cellcolor[HTML]{D1F1DA}SE & \cellcolor[HTML]{D1F1DA}z    & \cellcolor[HTML]{D1F1DA}p \\
\cellcolor[HTML]{F2F2F2}R & \cellcolor[HTML]{F2F2F2}L dlPFC  & 0.88                      & \textbf{\textless{}1E-3}  & \textbf{3.34\%}                         & \textbf{0.007}             & \textbf{4.03}             & \textbf{\textless{}1E-3}     & \textbf{12.58\%}                         & \textbf{0.015}             & \textbf{3.69}                & \textbf{\textless{}1E-3}  \\
\cellcolor[HTML]{F2F2F2}  & \cellcolor[HTML]{F2F2F2}L preSMA & 0.93                      & \textbf{\textless{}1E-3}  & \cellcolor[HTML]{FFFFFF}\textbf{5.06\%} & \textbf{0.006}             & \textbf{5.16}             & \textbf{\textless{}1E-3}     & \cellcolor[HTML]{FFFFFF}\textbf{8.11\%}  & \textbf{0.007}             & \textbf{5.13}                & \textbf{\textless{}1E-3}  \\
\cellcolor[HTML]{F2F2F2}  & \cellcolor[HTML]{F2F2F2}L cross  & 0.87                      & \textbf{\textless{}1E-3}  & \cellcolor[HTML]{FFFFFF}\textbf{5.49\%} & \textbf{0.799}             & \textbf{4.72}             & \textbf{\textless{}1E-3}     & \cellcolor[HTML]{FFFFFF}\textbf{12.59\%} & \textbf{0.014}             & \textbf{4.42}                & \textbf{\textless{}1E-3}  \\
\cellcolor[HTML]{F2F2F2}  & \cellcolor[HTML]{F2F2F2}R dlPFC  & 0.68                      & \textbf{\textless{}1E-3}  & \cellcolor[HTML]{FFFFFF}\textbf{2.24\%} & \textbf{0.007}             & \textbf{3.05}             & \textbf{0.00}                & \cellcolor[HTML]{FFFFFF}\textbf{12.53\%} & \textbf{0.016}             & \textbf{3.83}                & \textbf{\textless{}1E-3}  \\
\cellcolor[HTML]{F2F2F2}  & \cellcolor[HTML]{F2F2F2}R preSMA & 0.95                      & \textbf{\textless{}1E-3}  & \cellcolor[HTML]{FFFFFF}\textbf{5.24\%} & \textbf{0.005}             & \textbf{5.16}             & \textbf{\textless{}1E-3}     & \cellcolor[HTML]{FFFFFF}\textbf{7.46\%}  & \textbf{0.008}             & \textbf{4.77}                & \textbf{\textless{}1E-3}  \\
\cellcolor[HTML]{F2F2F2}  & \cellcolor[HTML]{F2F2F2}R cross  & 0.86                      & \textbf{\textless{}1E-3}  & \cellcolor[HTML]{FFFFFF}\textbf{6.95\%} & \textbf{1.269}             & \textbf{4.50}             & \textbf{\textless{}1E-3}     & \cellcolor[HTML]{FFFFFF}\textbf{17.00\%} & \textbf{0.022}             & \textbf{4.06}                & \textbf{\textless{}1E-3}  \\
\cellcolor[HTML]{F2F2F2}S & \cellcolor[HTML]{F2F2F2}L dlPFC  & 0.98                      & 0.075                     & \cellcolor[HTML]{FFFFFF}1.47\%          & 0.010                      & 0.70                      & 0.48                         & \cellcolor[HTML]{FFFFFF}\textbf{7.34\%}  & \textbf{0.009}             & \textbf{3.96}                & \textbf{\textless{}1E-3}  \\
\cellcolor[HTML]{F2F2F2}  & \cellcolor[HTML]{F2F2F2}L preSMA & 0.99                      & 0.315                     & \cellcolor[HTML]{FFFFFF}1.91\%          & 0.010                      & 1.07                      & 0.29                         & \textbf{8.83\%}                          & \textbf{0.011}             & \textbf{4.03}                & \textbf{\textless{}1E-3}  \\
\cellcolor[HTML]{F2F2F2}  & \cellcolor[HTML]{F2F2F2}L cross  & 0.98                      & 0.171                     & 1.47\%                                  & 1.108                      & 0.74                      & \cellcolor[HTML]{FFFFFF}0.46 & \textbf{8.59\%}                          & \textbf{0.011}             & \textbf{4.58}                & \textbf{\textless{}1E-3}  \\
\cellcolor[HTML]{F2F2F2}  & \cellcolor[HTML]{F2F2F2}R dlPFC  & 0.99                      & 0.624                     & 1.12\%                                  & 0.003                      & 1.96                      & 0.05                         & \textbf{2.02\%}                          & \textbf{0.004}             & \textbf{2.08}                & \textbf{0.005}            \\
\cellcolor[HTML]{F2F2F2}  & \cellcolor[HTML]{F2F2F2}R preSMA & 0.98                      & 0.359                     & 1.38\%                                  & 0.006                      & 1.41                      & 0.16                         & \textbf{3.99\%}                          & \textbf{0.006}             & \textbf{3.39}                & \textbf{\textless{}1E-3}  \\
\cellcolor[HTML]{F2F2F2}  & \cellcolor[HTML]{F2F2F2}R cross  & 0.99                      & 0.670                     & -1.40\%                                 & 0.006                      & -1.87                     & \cellcolor[HTML]{FFFFFF}0.06 & 2.14\%                                   & 0.008                      & \cellcolor[HTML]{FFFFFF}2.11 & 0.035                    
\end{tabular}
\caption{Tests for beta ($[15 - 30]$ Hz) power for monkeys R and S, for spectra and cross-spectra in the left (L) and right (R) hemisphere. MPC = Mean Percentage Change; SE = Standard Error. Values in bold indicate significance at a Bonferroni-corrected level of 0.0063}\label{tab:betaband}
\end{table}

\end{document}